\pdfoutput=1

\documentclass[11pt]{article}

\usepackage{ACL2023}

\usepackage{times}
\usepackage{latexsym}
\usepackage{caption}
\usepackage{subcaption}
\usepackage{algpseudocode}
\usepackage{graphicx}
\usepackage{algorithm}
\usepackage{amsfonts}
\usepackage{booktabs}
\usepackage{enumitem}
\usepackage[T1]{fontenc}

\usepackage[utf8]{inputenc}
\usepackage{dblfloatfix}
\usepackage{microtype}
\usepackage{mathtools}
\usepackage{wrapfig}
\usepackage{inconsolata}

\newcommand{\smallsec}[1]{\noindent {\bf #1.}}
\newcommand{\dataname}{ScreenRef}
\newcommand{\modelname}{SRR} 
\newcommand{\unamb}{Descriptive}
\title{Referring to Screen Texts with Voice Assistants}

\author{Shruti Bhargava, {\bf Anand Dhoot}, {\bf Ing-Marie Jonsson}, {\bf Hoang Long Nguyen},\\
{\bf Alkesh Patel}, {\bf Hong Yu}, {\bf Vincent Renkens}\\Apple Inc. \\
  \small \{shruti\_bhargava, adhoot, ingmarie, romanhoangnguyen\_long, alkesh.patel, hong\_yu, vrenkens\}@apple.com }

\begin{document}
\maketitle
\begin{abstract}
Voice assistants help users make phone calls, send messages, create events, navigate and do a lot more. However assistants  have limited capacity to understand their users' context. In this work, we aim to take a step in this direction. Our work dives into a new experience for users to refer to phone numbers, addresses, email addresses, urls, and dates on their phone screens. Our focus lies in reference understanding, which becomes particularly interesting when multiple similar texts are present on screen, similar to visual grounding. We collect a dataset and propose a lightweight general purpose model for this novel experience. Due to the high cost of consuming pixels directly, our system is designed to rely on the extracted text from the UI. Our model is modular, thus offering flexibility, improved interpretability, and efficient runtime memory utilization.
\end{abstract}

\section{Introduction}


With the advent of internet and smartphones, the world came to our fingertips. And with the emergence of voice assistants (VAs), everything became even more accessible. VAs have become pervasive in the smartphones as they offer natural means of communication to the user. They able a user to perform tasks faster with natural language instead of several taps, app switches, scrolls and typing. However, they are limited in their ability to understand the user's context.

\begin{figure}[bhp]{}
    \centering
\includegraphics[width=0.48\textwidth]{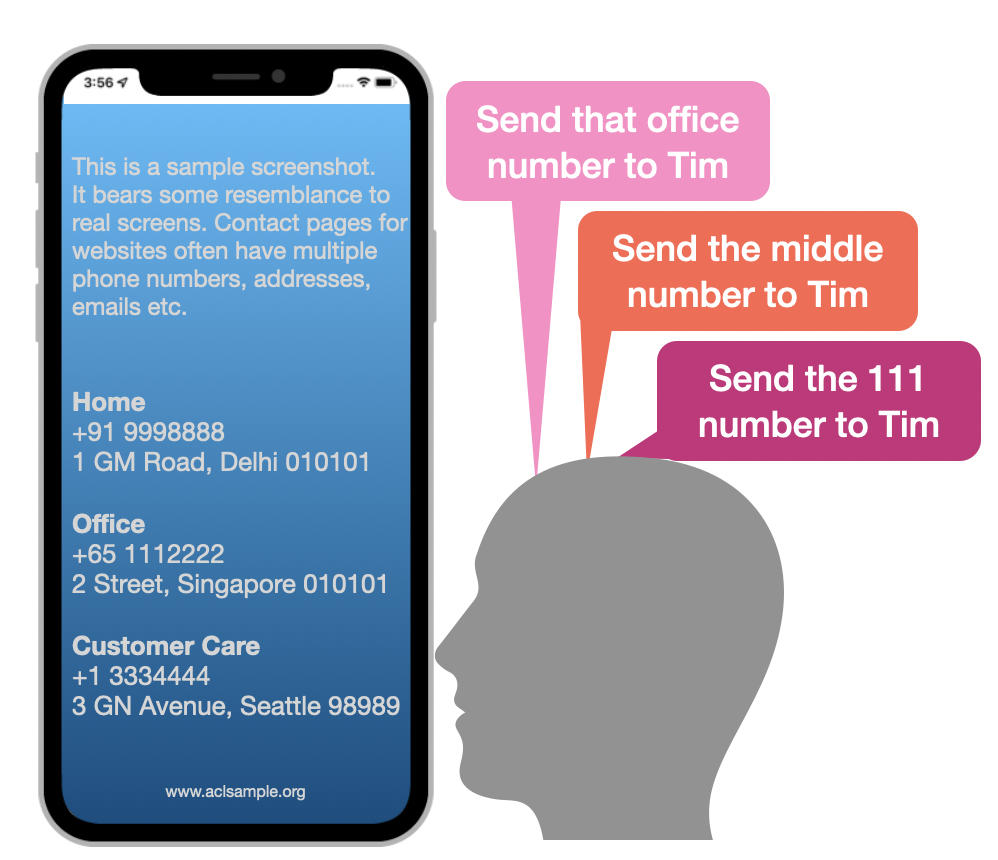}
    \caption{Suppose a user wants to share a number on their screen. We aim to support this in a natural and succinct way by enabling users to refer to screen elements in interactions with Voice Assistants.}
    \label{fig:intro}
\end{figure}

Let us look at an example. In Fig. \ref{fig:intro}, a user wants to share a number from a webpage to a friend. They might do either of the following:
\begin{itemize}[topsep=0pt, itemsep=0pt, parsep=0pt]
    \item memorize the number $\rightarrow$ go to messages $\rightarrow$ new message to friend $\rightarrow$ type the number from memory $\rightarrow$ send 
    \item select the number $\rightarrow$ copy $\rightarrow$ go to messages $\rightarrow$ new message to friend $\rightarrow$ paste $\rightarrow$ send
\end{itemize}

One solution might be that the user can read out the number to the VA. However reading out may be cumbersome and unnatural as this is not how one would communicate with a person standing next to them. Further, it may create unwarranted ASR errors, especially for texts like URLs or emails. Our work explores how to make this simpler by enabling users to refer to screen elements in requests made to voice assistants. References make conversations more natural and succinct, thus allowing the user to say: ``Send the middle number to Tim''.

We conduct a user study to explore how users would make requests involving screen elements. Participants are shown screenshots, each containing multiple entities of a category (eg. 3 phone numbers), and asked to type requests for a VA to act on one of them. The study reveals that a majority of users (57\%) prefer to use references like ``Send that office number to Tim' instead of repeating the full text.

For supporting such experiences, voice assistants need to resolve the references. In this work, we focus on such reference resolution. Specifically, we consider requests referring to phone numbers, addresses, email addresses, URLs, date/time. 
We choose these \textit{actionable text} since, $\sim$50\% of screen elements are texts ~\cite{zhang2021screen}, and these categories are commonly acted upon. Users often call or message numbers, share contact details, navigate to addresses.

To understand and evaluate the task, we collect \dataname, a dataset of 14k requests, with references to \textit{actionable text} or entities. {\dataname} contains two collections. First, \textit{{\unamb} Data}, which is based off screens with multiple similar entities to get descriptive references like `call the Apple Business Manager number'.  This is similar to how visual grounding datasets~\cite{kazemzadeh2014referitgame, mao2016generation} focus on images with multiple objects of the same category to get challenging references. Second, \textit{Category level Data}, includes simpler references to a category without disambiguating within a category, eg. `call this number'. These are described in Sec. \ref{sec:dataset}.

With the purpose of deploying in real world, it is critical to design solutions with low latency. To this end, we design Screen Reference Resolver, or \modelname, a modular attention based architecture for reference understanding. We focus on privacy, hence our model is lightweight and executable on-device. Our network re-uses existing signals available from upstream including request embedding, and text scraped from the UI. We also discuss a heuristic-based baseline (designed for quick prototyping).  

Overall, our main contributions are: 
\begin{enumerate}[topsep=4pt, itemsep=0pt]
\item We explore a novel experience for Voice Assistant users to execute tasks on \textit{actionable text} on the phone screen by using references.
\item We conduct a user study to analyse users' interactions with entities on screen. This reveals interesting insights about usage of references.
\item We design efficient data collection schemes for collecting requests with references to \textit{actionable text} on screen and collect a dataset.
\item To understand references to entities on screen, we propose a heuristic-based baseline and a modular attention-based network, \modelname. The model has a small memory footprint, low latency, can run on device, and drastically boosts performance compared to the baseline. 
\end{enumerate}

\section{Related Work}

\smallsec{Grounding to UI elements} Past works have explored mapping natural language commands to UI elements for Chrome web pages \cite{pasupat-etal-2018-mapping},  grounding executable actions for UI navigation \cite{li-etal-2020-mapping} and user interaction \cite{xu-etal-2021-grounding}. These works primarily focus on navigational commands, thus target buttons, links and input boxes. Our goal is to explore screen referencing capability for common VA tasks, thus we target `actionable text entities' like phone numbers. \citet{hsiao2022screenqa} propose ScreenQA with questions about UI elements including text, which could also benefit from UI grounding. \citet{wang2022enabling, rozanova2021grounding} investigate LLM abilities for UI grounding. \citet{li2021vut, li2022spotlight} use vision and language transformers for the task. However, we only use the screen texts and no pixels directly. Our solution design focuses on low latency, less memory and privacy-preserved inference that can be run on device.

\smallsec{Voice assistants and Multimodal Interactions} The power of replacing multiple low-level actions by natural language has been explored for webpage designing ~\cite{kim2022}, image editing \cite{laput2013pixeltone}. Users use VAs for controlling screen content, particularly the visually challenged \cite{vtyurina2019}.  ~\citet{ljungholmvoice, luger2016like} discuss how lack of context understanding makes VA usage unnatural. ~\citet{bolt1980} employed a point-and-speak approach for desktops. Prior works have explored tracking user gaze for multimodal interactions~\cite{drewes2007}, for digital screens \cite{hutchinson1989human, mardanbegi2011mobile} as well as for external, real-world objects \cite{mayer2020}. In this work, we explore using language to reduce the low-level actions needed to interact with certain text categories on phone screens and thereby increase the context understanding of VAs. 


\smallsec{Grounding to objects and text in open scenes} A related task to ours is visual grounding~\cite{kazemzadeh2014referitgame, mao2016generation, yu2018mattnet}, resolving references to physical objects in scenes. The physical form and semantics of text is much different, resulting in different reference forms. \citet{rong2019unambiguous, rong2017unambiguous} look at references to text in scenes. However, a lot of their references are of the form `the text on \ldots', thus grounding requires less knowledge of text and more of physical objects. Also, the major challenge in open scenes is text localisation and recognition, which is much simpler on phone screens. On the other hand, screens are challenging as they contain a lot more text. TextVQA, from \citet{singh2019towards, biten2019scene} could utilise grounding to text, but doesn't contain labels for this. Lastly, none of these works cater to task-oriented dialog for text on screen, which is our primary focus. 

 \section{User Study}
To understand how users would make requests about entities on the screen, we conduct a user study on Pollfish~\cite{pollfish}. We use 4 screens for each \textit{actionable text category} (phone numbers, emails, addresses, and URLs), and each screen has 3 instances of the category (eg. 4 screens with 3 phone numbers each). A total of 300 participants are selected from across US, balanced for VA usage, gender and English as first or other language. Overall, 4800 typed requests were collected. 

\begin{figure}
        \centering
        \includegraphics[width=0.48\textwidth]{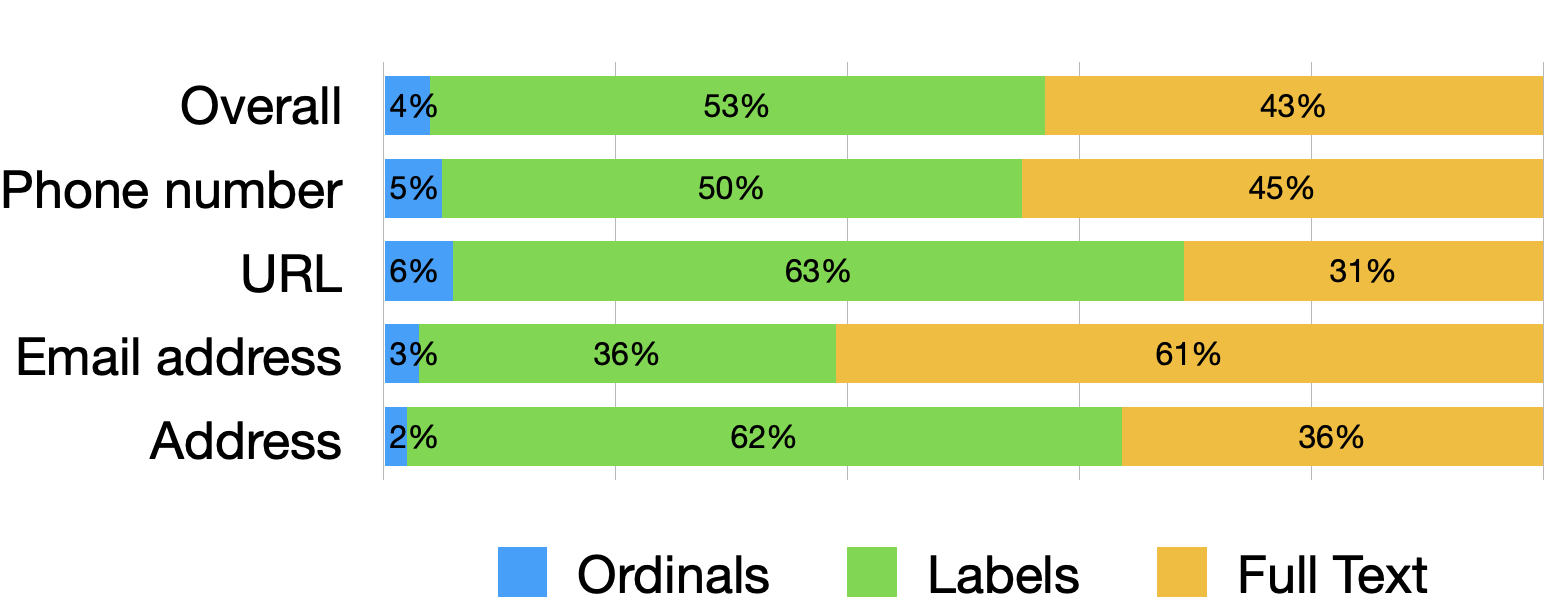}
        \caption{Distribution of request types in the user study. References using labels i.e. text within or around entity are most common, followed by repeating full text. For entities like addresses and URLs, repeating full text may be cumbersome, hence references are more common.}
    \label{fig:ref_types_study_distri}
\end{figure}
The responses were reviewed by two researchers. Using heuristics, three common types of requests surfaced:  1. Full Text: ``call 1-866-902-7144'' 2. Labels: using text other than full entity text ``directions to the one in Portland'' 3. Ordinals: ``send the third email address''. The data shows a heavy preference towards the first two (Fig. \ref{fig:ref_types_study_distri}). Intuitively, when browsing information, the eyes are often scanning for a topic of interest. For instance, “I need to call support”, explaining the label based requests. Our hypothesis for the high use of full text is that they didn't want to rely on VA's ability to understand the context. Within references, using the text in or around the entity is common and the position is used sometimes. Note that our study was performed on a limited set of users and for a limited set of screens, but we uncover interesting patterns on how users might request actions on screen texts. It is important to keep in mind that speaking full texts could be cumbersome, unnatural and have speech recognition errors, especially for entities like URLs.

\section{Task and Dataset}
\label{sec:dataset}
Given a screen S (with OCR texts $t$), text entities  $e_1,...,e_k$ and a request $r$, the task is to select the entity(entities) $e \in {e_1, . . . , e_k}$ being referred to in $r$.

We collect \dataname, a collection of requests to Voice Assistants with references to actionable text categories on screen (phone number, address, email address, URL, date/time). Due to privacy concerns on sharing a dataset with extracted phone numbers/emails from web pages, we are unable to share the dataset but we discuss the collection protocol in detail (see samples in Fig.~\ref{fig:data_sample} and annotation guidelines in Appendix \ref{sec:appendix}). We collect full requests, not just reference phrases, since words outside of the explicit reference phrase may hint at the targeted entity. For instance, \textit{call this} has the reference \textit{this} which is ambiguous, however the request can be understood as referring to a phone number. 20 annotators are recruited for our data collections.

\begin{figure}
    \begin{center}
\includegraphics[width=0.48\textwidth]{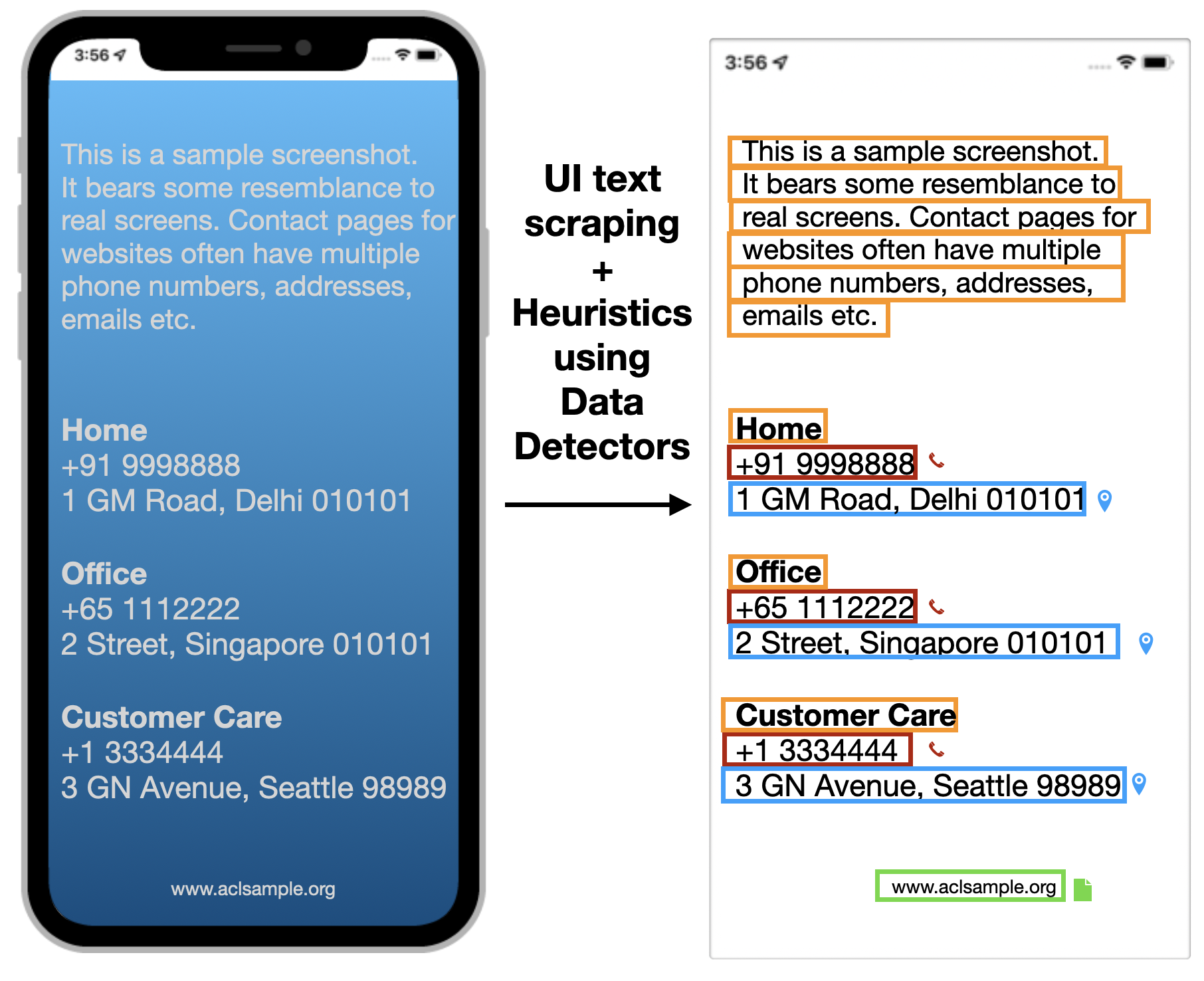}
    \end{center}
    \caption[]{Screen processing is done by upstream systems using data detectors to get entity categories. We get texts with their location, and texts classified as phone, email, address, date/time, URL that form the candidate entities. Example entity: 
        [\textit{text:} +91 9998888, \textit{location:} [0.04, 0.36, 0.4, 0.03], \textit{category:} Phone Number]. These are the inputs to our grounding system.}
    \label{fig:screen}
\end{figure}

We first started with a simple collection protocol. After extracting entities of our interest using data detectors for a list of web pages, we show a web page screen with one highlighted entity and ask graders to provide a request referring to that entity. This would get us a dataset of requests referring to screen entities and their referred entity. However, this ran into major issues. First, annotators would often miss other similar entities on the screen and provide requests which are ambiguous, eg. ``call this number'', when there is more than 1 number on the screen and only 1 of them is highlighted, thereby resulting in an incorrect sample for the resolution task. Second, there were a large number of duplicate requests ($>$40\%). This may happen due to several screens may have one entity and thereby annotators may use simple references. Other issues included the need of screens with entities to collect any data, lack of representation of different reference types within the collected data and lack of awareness of ambiguous requests. 

The quality and efficiency concerns led us to develop a new protocol in the form of descriptive and category level data collections. Within descriptive collection, we use a similar screen based collection technique. However we restrict to screens with more than one instance of a category in order to collect challenging and diverse requests (similar to visual grounding datasets like RefCOCO). Alongside the target entity, we highlight all entities of that category to reduce chances of erroneous ambiguous requests. Within category level collection, we do not use screens and the focus is on unique diverse requests with simple references. This split addresses the issues described above leading to more efficient collection and better quality datasets.

\begin{figure}[t]
    \centering
    \includegraphics[width=0.47\textwidth]{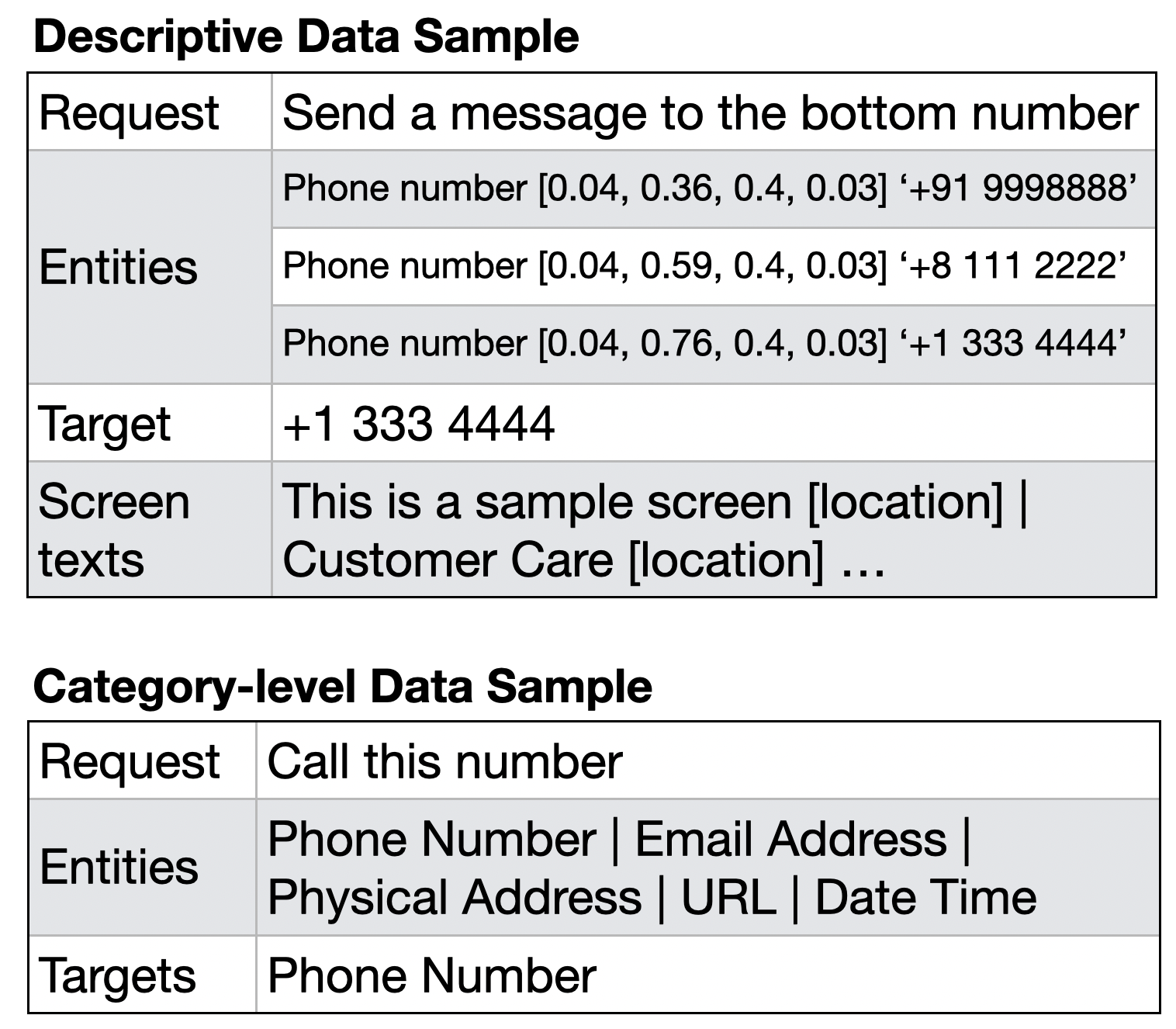}
    \caption{Samples from \dataname. {\unamb} data is collected using screens and has entities and texts from the screen. Category-level data is collected without screens and has an entity pool containing one dummy entity from each of our scoped categories.}
    \label{fig:data_sample}
\end{figure}

\smallsec{Descriptive Data Collection}
Though we aim to support references on all apps on phone, this collection is carried out with web pages due to their varied layouts and ease of access. For a list of top visited web pages, we extract texts by UI scraping and get text categories using data detectors. This is similar to running object detectors in open scenes. In order to get challenging references, we only keep screens with more than one entity from a category (eg. 2 URLs).\\
One entity on the screen is highlighted as target and users are asked to provide requests for that (Fig. \ref{fig:unamb_collection}). Guidelines provided in \ref{descriptive_guidelines}. For quality check, we run a verification to confirm the requests are unambiguous: three independent annotators are shown the screen and the collected request and asked which entity from the screen is the request referring to. Samples where at least 2/3 annotators agree are kept, leading to $\sim$6\% data drop.

\smallsec{Category-level Data Collection}
This collection targets \textit{simple references} for a category (``phone number - Call that number'', ``URL - Open it''). During screen mining, we observe that a lot of the screens have only one entity from a category. In these cases, users may prefer succinct simple references instead of descriptive ones. Note that these references do not use the screen layout. Hence, we design this collection independent of screens. This gives a simpler collection scheme that allows us to scale to new categories and/or locales more quickly, with reduced time and cost.\\
We show a category and ask annotators to give requests, assuming that entity is on their screen. The collection is carried out on shared spreadsheets, one sheet per category (Fig. \ref{fig:cat_collection}) in order to avoid duplicate requests across annotators. Annotators are given automatic instant feedback by COUNT\_UNIQUE to encourage variations. Through pilot annotation projects, we recognize several constraints to ensure that the uniqueness is not from spurious modifications, which are also added to the guidelines. Detailed grading guidelines provided in \ref{category_guidelines}. For verification, 3 independent annotators are shown a request and asked to mark \textit{all} categories it could refer to. This also gives annotated multi-label samples i.e. requests that are category ambiguous: ``take me there'' could be referring to a \textit{URL} or an \textit{address}. Requests with majority agreement are kept. After the requests are collected, dummy entities, one of each scoped category, are added to each request to form a data sample. In a way, this makes the dataset more complete and challenging than real screens which may include only a subset of the entity categories. (Fig. \ref{fig:data_sample}).
 \begin{table}[!b] 
 \resizebox{\columnwidth}{!}{
  \begin{tabular}{lcccc}
    \toprule
    &\multicolumn{2}{c}{Category-level} & \multicolumn{2}{c}{Descriptive}\\
     &Train&Test&Train&Test\\
    \midrule
    Total requests & 4137 & 486 & 7993 & 1082\\
    Unique requests & 4123 & 486 & 6520 & 957 \\
    Multilabel & 934 & 126 & 0 & 0\\
    Tokens per request & 7.78 & 7.95 & 7.46 & 7.65\\
    Tokens per reference & 2.09 & 2.06 & 4.25 & 4.31\\
    Screenshots & - & - & 336 & 42 \\
  \bottomrule
\end{tabular}
}
\caption{Statistics for all requests in {\dataname}}
\label{tab:freq}
\end{table}

\begin{figure*}[btp]
     \centering
     \begin{subfigure}[b]{0.33\textwidth}
        \centering
\includegraphics[width=\textwidth]{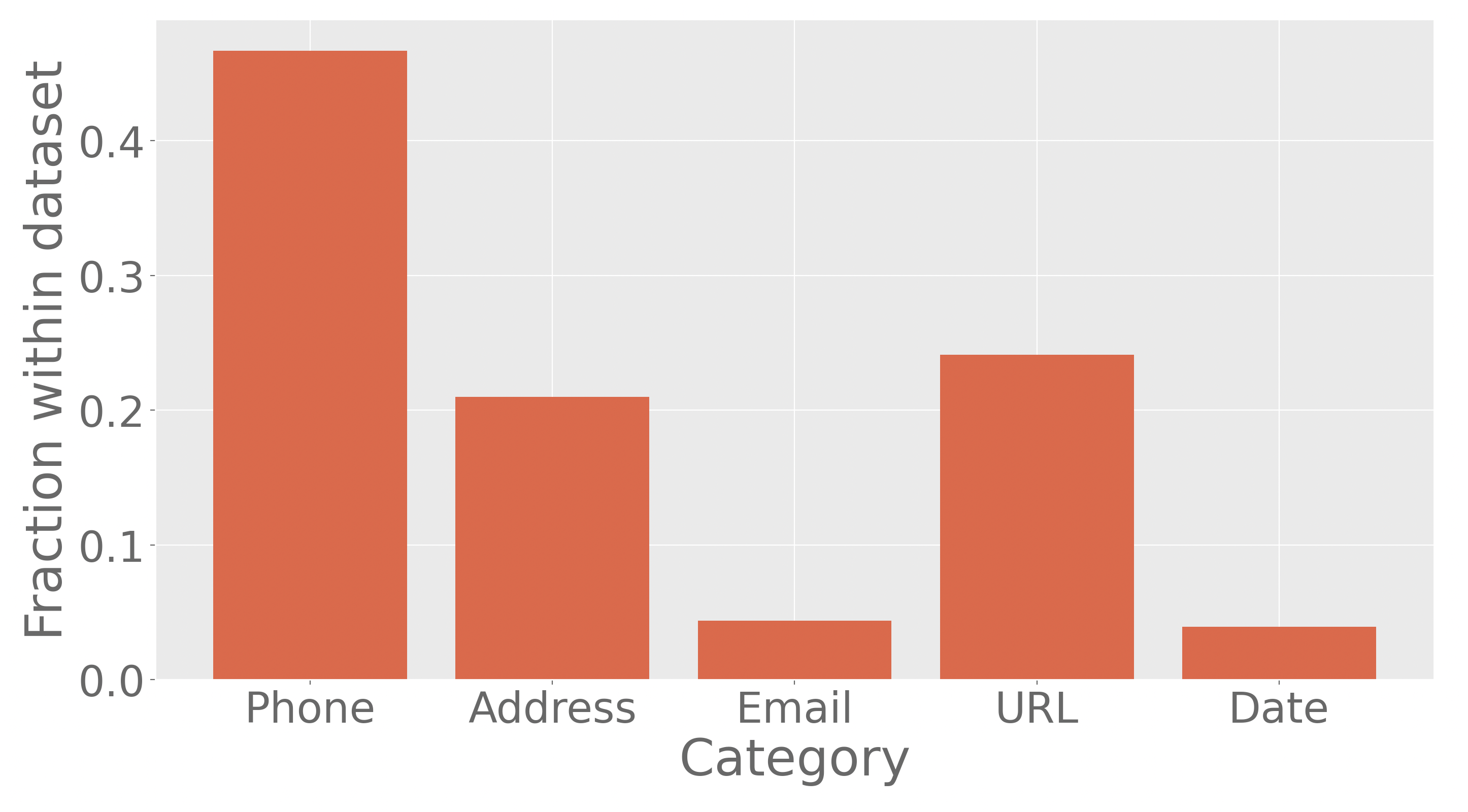}
        \caption{Requests across categories.}
        \label{fig:entity_distribution}
     \end{subfigure}
     \hfill
     \begin{subfigure}[b]{0.32\textwidth}
        \centering
        \includegraphics[width=\textwidth]{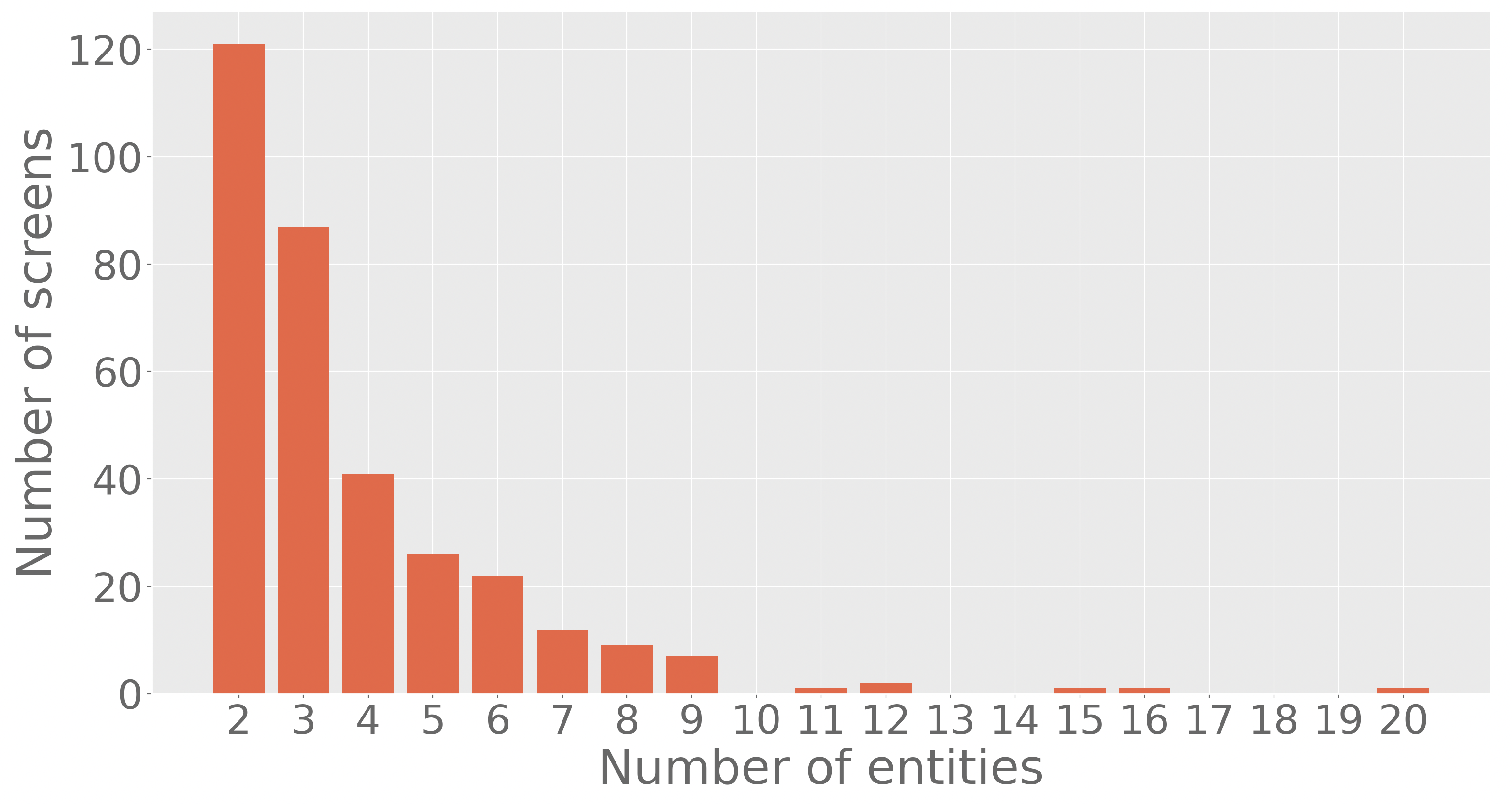}
        \caption{\# of entities on screens.}
        \label{fig:screen_entity_distribution}
     \end{subfigure}
     \hfill
     \begin{subfigure}[b]{0.32\textwidth}
        \centering
        \includegraphics[width=\textwidth]{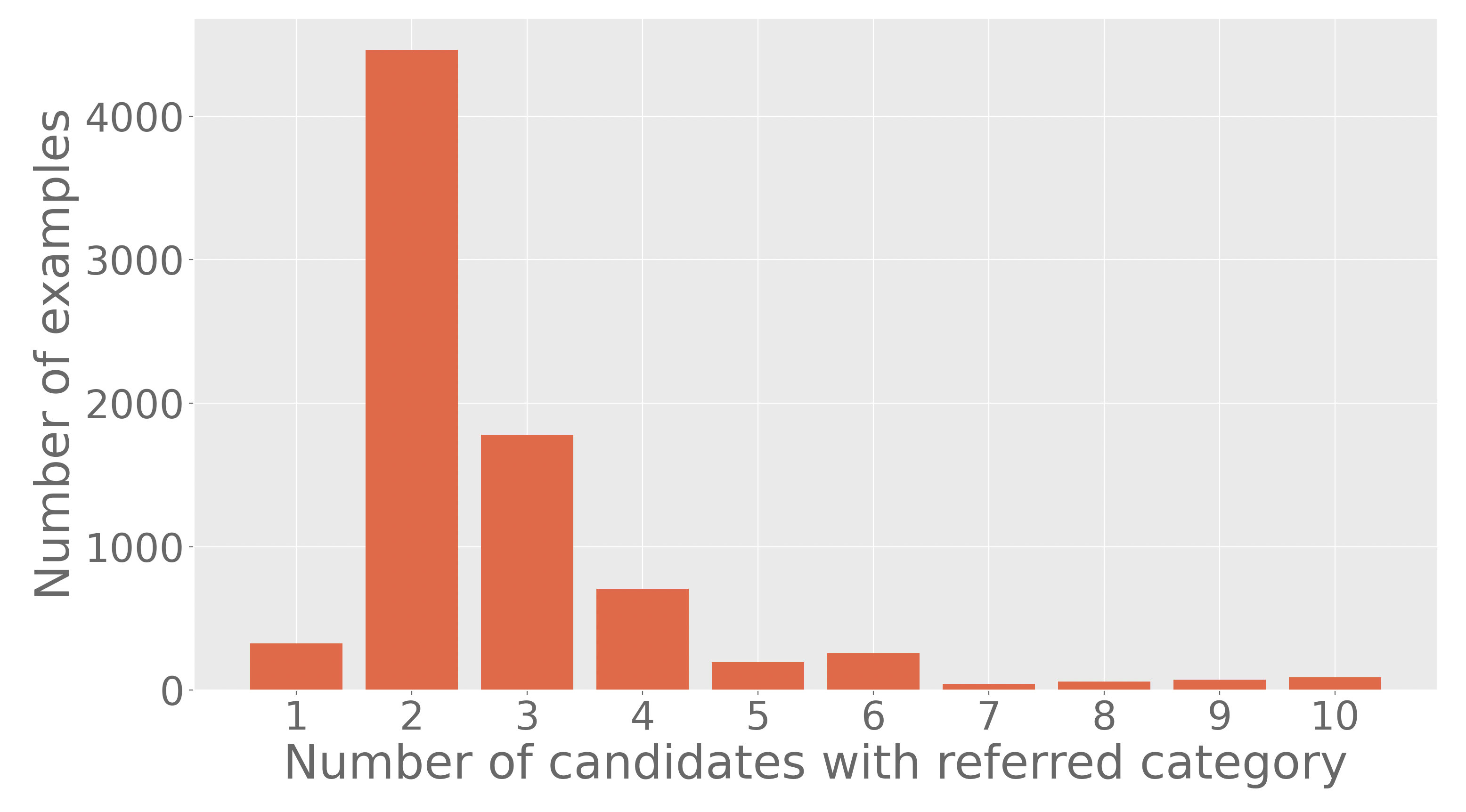}
        \caption{\# of entities with same category as gt.}
        \label{fig:targeted_entity_distribution}
     \end{subfigure}
        \caption{Histogram of various factors in the {\unamb} Data}
        \label{fig:histograms}
\end{figure*}


\section{Models}
\subsection{Heuristic-based Baseline}



This is designed for quick prototyping and development without much training data. We define a set of hand-crafted rules using keywords from a subset of the training data.  The rules are applied in sequence: 

\begin{enumerate}[topsep=4pt, itemsep=1pt, parsep=1pt,leftmargin=5.5mm]
    \item \smallsec{Phrase-match} Look for synonyms or verbs or apps in the request that indicate the target category (like `number', `call' indicate \textit{phone number},  `navigate', `maps' indicate \textit{address}). 
    \item \smallsec{Location-match} Regex match to find positional or ordinal reference in request, sort candidates by coordinates and pick the entity at the mentioned position.
    \item \smallsec{Label-match} Locate the text on screen that has maximum match to the request using a set of string matching features like word overlap (after removing stopwords). Pick the entity closest to this text.
    \item If none of the entities are selected above (like ``Share this''), score all entities identically.
\end{enumerate}

\vspace{-2mm}
\subsection{Screen Reference Resolver}
\begin{figure}[tbp]{}
    \centering
\includegraphics[width=0.5\textwidth]{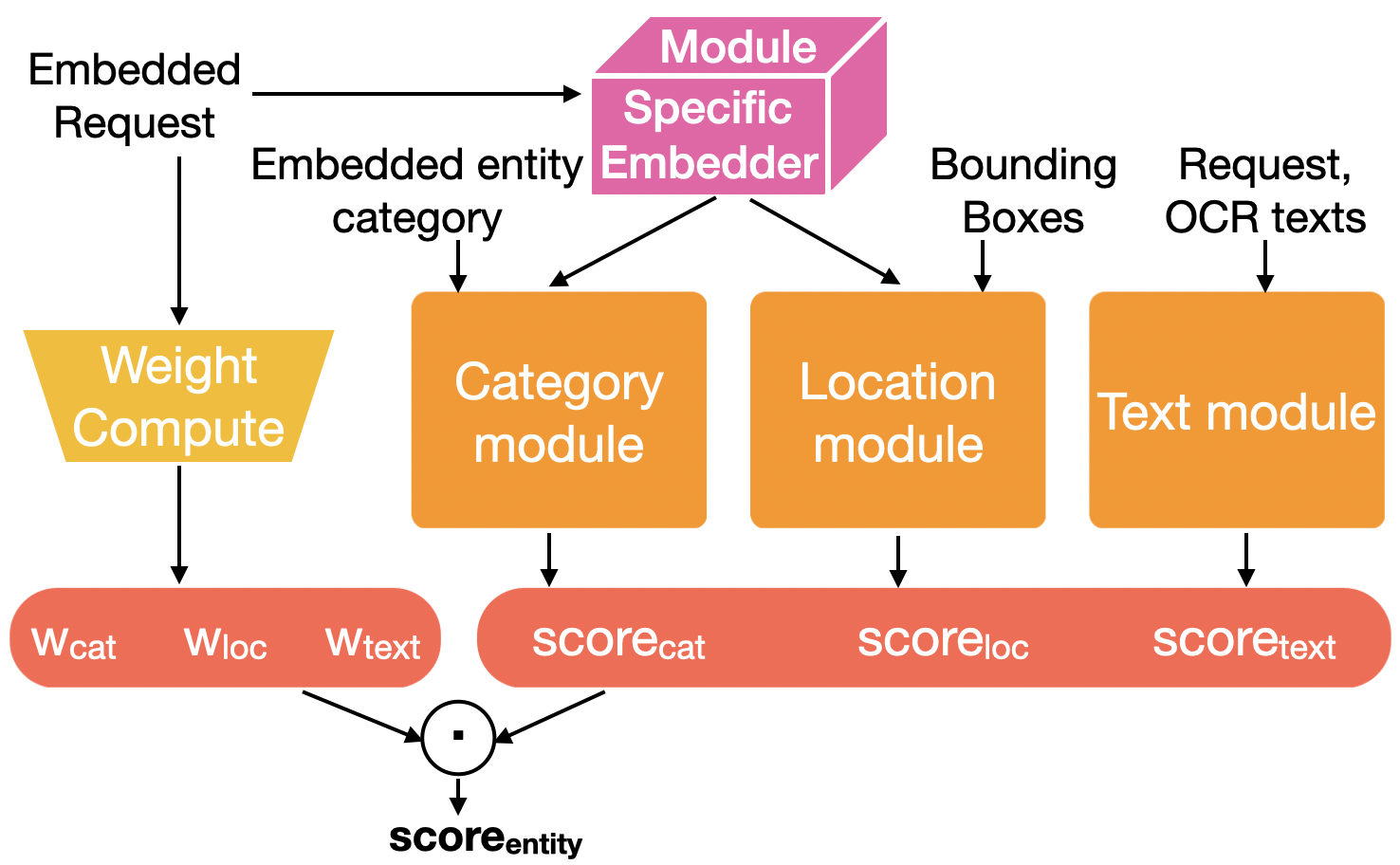}
    \caption{Architecture for the Screen Reference Resolver. It uses the embedded request, embedded entity category, location features and text matching features to predict a matching score for the entity and the request.}
    \label{fig:model}
\end{figure}

We design \modelname, a modular attention-based network for resolving references (Fig. \ref{fig:model}). Inspired by MattNet \cite{yu2018mattnet}, the model contains 3 modules, each of which use a subset of signals from entities, use soft attention to attend to relevant tokens of the request, and compute relevance scores for each entity with the request. We focus on two key dimensions crucial for deploying in an industrial setup- first, memory footprint; second, reusing the existing components in the pipeline. 

We re-use the request token embeddings generated by the upstream embedder (like Bert \cite{devlin2018bert}) and the text categories recognized by upstream. The embedded request passes through the \textit{weight compute block}, an MLP followed by softmax, that predicts weights for each module. A request like ``call the top phone number'' could give high weight to location and category modules, while ``call the one in Palo Alto'' could give higher weight to the text and category modules. Embedded tokens also go to the \textit{module-specific embedder} where soft attention is applied on the token embeddings to get embeddings independently for the category and location modules. For ``call the top phone number'', category module could attend more to `call' and `phone number', while location to `top'. Modules produce scores by fusing entity features with these embeddings. Module scores are combined using the module weights to get the final score for each entity. Specifically, the final score is $ w_{cat} \times s_{cat} + w_{loc} \times s_{loc} + w_{text} \times s_{text}$. Let us understand the three modules.

\smallsec{Category module} Entity categories (phone number, URL etc) are embedded using the same embedder as the request. These are pre-computed for all categories. During runtime, given an entity and a request, the embedding for the entity category is loaded and matched with the request embedding from the \textit{module specific embedder}. Both are passed through separate MLP blocks, followed by an inner-product to compute the matching score.

\smallsec{Location module} This takes in bounding boxes of the entity and of other entities of the same category (similar to \cite{yu2018mattnet}). Bounding boxes of entities $[x,y,w,h]$ are normalized by $K = max(I_{width}, I_{height})$, preserving the aspect ratio and featurised as:
\begin{gather*}
    \left[ \frac{x}{K}, \frac{y}{K}, \frac{x+w}{K}, \frac{y+h}{K}, \frac{w*h}{K^2} \right]
\end{gather*}These features are concatenated and passed through an MLP. Embedding from the \textit{module specific embedder} is passed through a separate MLP, and lastly an inner product gives the module score.

\smallsec{Text module} We do not embed the screen texts but instead use string matching features. This choice is made for three main reasons. First, we observe in the user study users typically use the text and not synonyms of the text present on screen when making references. Second, our entities of interest, like numbers, emails, URLs make little sense to embed due to their content and presence of OOVs. Third, screens can have a large number of texts. Embedding so many texts in run time could cause compute overhead. Hence, instead of embedding, we utilize the texts by designing simple features like: is the text fully contained in the request, word overlap after removing stopwords, digit overlap. Along with matching the request to the entity text, we also match with the entity's neighboring texts (sorted by distance). All features are concatenated and passed through an MLP to get the module score.

Since the Category-level data has multi-label instances (eg. `take me there' could refer to a URL or an address), we use a threshold (a hyperparameter obtained from fine-tuning on val data as 0.7) to get final predictions. We add intermediate supervision on the module weights by annotating $\sim$500 samples each for ordinal references (labelled for high weight to location module), references using visible text (text module), and simple reference (category module). 

Modular nature offers memory efficiency, giving an option to skip running some modules which get very low weights for a request. It also provides flexibility for varied reference resolution use cases, eg. scenarios with only entity categories available. Note that {\modelname} is only 1MB in size, does not need access to DOM or view hierarchy and hence can work on any screen, in fact any context where recognized texts are available, including documents. 
\begin{table}[!b]
    \begin{tabular}{llccc}

\toprule
  Dataset     & Model &  Top-1 Err. & EM \\
\midrule
        Category- &Heuristic & 6.5& 87.5  \\

    level&SRR  & \textbf{1.1} & \textbf{89.9} \\ 
    &Cat. Oracle & 0.0 & 100\\
    & No text Oracle & 0.0 & 100\\
\midrule
    {\unamb}  & Heuristic  &  25.0 & 74.2\\

    &SRR &  \textbf{14.2}  & \textbf{78.7}\\
    &Cat. Oracle  & 54.0 & 0.0\\
    & No text Oracle & 32.6 & 45.5\\
\bottomrule
    \end{tabular}
    \captionof{table}{Top-1 Error and Exact Match accuracy of various systems on \dataname. \modelname~ reduces the relative top-1 error by 83\% on category-level data and 43\% on {\unamb} data compared to the heuristic baseline. Category Oracle predicts all entities of the true category. Exact Match going from 100 to 0 and top-1 error from 0 to 55 between the two subsets shows how they differ by design. No text Oracle knows all simple and ordinal references but not the text values.}
    \label{tab:baselines}
\end{table}

\section{Results}

\smallsec{Experimental Setup} Data is split into train/val/test in 80/10/10 ratio. To avoid data leak in {\unamb} data, we split the data by screens, thus all requests for a screen are in one set. Table \ref{tab:freq} summarizes the overall statistics for the dataset. We randomly pick a negative sample for each positive sample and use binary cross entropy loss and Adam optimizer with an initial learning rate of $4 \times 10^{-4}$. 

\smallsec{Metrics} We use two metrics to measure performance. First, \texttt{exact match accuracy} indicates whether the predicted entities, after applying the threshold, exactly match the true entities (if an additional entity crosses the threshold or one of the true entities doesn't, exact match is 0). Second, \texttt{top-1 error} indicates whether the entity with the highest score, regardless of threshold, is in the ground truth entities. This is useful as often only the top prediction is used by downstream.

\begin{table}
\centering
{
\begin{tabular}{cccc}
     \toprule
     \multicolumn{3}{c}{Modules in \modelname} & \\
    Category & Location & Text & Top-1 Error  \\
     \midrule
     $\checkmark$ & $\checkmark$ & $\checkmark$ & \textbf{14.2} \\ 
      $\times$ & $\checkmark$ & $\checkmark$& 31.2 \\
       $\checkmark$& $\checkmark$& $\times$ & 33.7 \\
       $\checkmark$ & $\times$ & $\checkmark$ & 35.3\\
       $\times$ & $\checkmark$ & $\times$ & 49.7\\
        $\times$ & $\times$ & $\checkmark$ & 51.5\\
       $\checkmark$ & $\times$ & $\times$  & 54.9\\ 
        \bottomrule
    \end{tabular}
    }
    \captionof{table}{Ablation Study results for the different modules in {\modelname} namely  category, location and text modules. Top-1 Error on the {\unamb} data is reported. The observed loss in performance across all subsets underscores that all modules are critical for achieving high performance.}
    \label{tab:ablation}
    \end{table}
    
\smallsec{Results} We summarize the results in Table \ref{tab:baselines}. Observe that the performance on Category data is higher than on the {\unamb} data, 
indicating the challenging nature of the latter. {\modelname} reduces the relative top-1 error by 43\% on {\unamb} and 83\%  on Category-level data compared to the baseline. The oracles know the true category hence get perfect results on category-level data. Their low performance on {\unamb} reflects the importance of all inputs, particularly screen texts. We carry out an ablation study on the model (Fig. \ref{tab:ablation}). It shows that each attribute and thus each module is critical in understanding the references.

\section{Conclusion}
We explore a new user experience of executing actions on screen elements with Voice Assistants. To make interactions more natural, we explore the use of references. An important decision was what UI elements to support. We decided to use texts that are most commonly used for task oriented dialogue and, commonly present on phone screens and easy to classify. We collected a dataset of requests and proposed solutions to understand references. This is a step towards making Voice Assistants more context aware, but there is a lot more context. We hope that our work will motivate further research towards this goal, and towards semantic visual text referencing.

\section*{Limitations}
Our work explores a dimension of context understanding by Voice Assistants but it is only a small step.  Firstly, we only consider 5 categories, while screens have a myriad of other texts and visual content. We do not include image context into our reference understanding models. But users could use them when formulating references to texts near them. Using image captions or some pixels would improve coverage. Our system leverages entities extracted by upstream and hence is bounded by the performance of that. Also our model evaluates each entity separately while there may be benefit in considering the entire screen holistically.

\section*{Ethics Statement}
This work aims at improving user experiences with voice assistants. By allowing users to refer to entities on screen, it reduces user friction and enables a smoother and more natural experience. No voice assistant usage log data was used and all requests were collected by recruited annotators.

\section*{Acknowledgements}
We would like to thank Hadas, Lucia and Kyanh for their help with the data annotations, revisions and data quality check; Sachin and Dhivya for help with data planning; Melis and Junhan for support in modelling experiments; as well as Lin and Murat for general direction.

\bibliography{anthology,paper}

\begin{thebibliography}{28}
\expandafter\ifx\csname natexlab\endcsname\relax\def\natexlab#1{#1}\fi

\bibitem[{pol()}]{pollfish}

\newblock {Pollfish}.
\newblock \url{https://www.pollfish.com}.

\bibitem[{Biten et~al.(2019)Biten, Tito, Mafla, Gomez, Rusiñol, Valveny,
  Jawahar, and Karatzas}]{biten2019scene}
Ali~Furkan Biten, Ruben Tito, Andres Mafla, Lluis Gomez, Marçal Rusiñol,
  Ernest Valveny, C.~V. Jawahar, and Dimosthenis Karatzas. 2019.
\newblock \href {https://doi.org/10.48550/ARXIV.1905.13648} {Scene text visual
  question answering}.

\bibitem[{Bolt(1980)}]{bolt1980}
Richard~A. Bolt. 1980.
\newblock \href {https://doi.org/10.1145/800250.807503} {“put-that-there”:
  Voice and gesture at the graphics interface}.
\newblock In \emph{Proceedings of the 7th Annual Conference on Computer
  Graphics and Interactive Techniques}, SIGGRAPH '80, page 262–270, New York,
  NY, USA. Association for Computing Machinery.

\bibitem[{Devlin et~al.(2018)Devlin, Chang, Lee, and
  Toutanova}]{devlin2018bert}
Jacob Devlin, Ming-Wei Chang, Kenton Lee, and Kristina Toutanova. 2018.
\newblock Bert: Pre-training of deep bidirectional transformers for language
  understanding.
\newblock \emph{arXiv preprint arXiv:1810.04805}.

\bibitem[{Drewes et~al.(2007)Drewes, De~Luca, and Schmidt}]{drewes2007}
Heiko Drewes, Alexander De~Luca, and Albrecht Schmidt. 2007.
\newblock \href {https://doi.org/10.1145/1378063.1378122} {Eye-gaze interaction
  for mobile phones}.
\newblock In \emph{Proceedings of the 4th International Conference on Mobile
  Technology, Applications, and Systems and the 1st International Symposium on
  Computer Human Interaction in Mobile Technology}, Mobility '07, page
  364–371, New York, NY, USA. Association for Computing Machinery.

\bibitem[{Hsiao et~al.(2022)Hsiao, Zubach, Wang et~al.}]{hsiao2022screenqa}
Yu-Chung Hsiao, Fedir Zubach, Maria Wang, et~al. 2022.
\newblock Screenqa: Large-scale question-answer pairs over mobile app
  screenshots.
\newblock \emph{arXiv preprint arXiv:2209.08199}.

\bibitem[{Hutchinson et~al.(1989)Hutchinson, White, Martin, Reichert, and
  Frey}]{hutchinson1989human}
Thomas~E Hutchinson, K~Preston White, Worthy~N Martin, Kelly~C Reichert, and
  Lisa~A Frey. 1989.
\newblock Human-computer interaction using eye-gaze input.
\newblock \emph{IEEE Transactions on systems, man, and cybernetics},
  19(6):1527--1534.

\bibitem[{Kazemzadeh et~al.(2014)Kazemzadeh, Ordonez, Matten, and
  Berg}]{kazemzadeh2014referitgame}
Sahar Kazemzadeh, Vicente Ordonez, Mark Matten, and Tamara Berg. 2014.
\newblock Referitgame: Referring to objects in photographs of natural scenes.
\newblock In \emph{Proceedings of the 2014 conference on empirical methods in
  natural language processing (EMNLP)}, pages 787--798.

\bibitem[{Kim et~al.(2022)Kim, Choi, Choi, and Kim}]{kim2022}
Tae~Soo Kim, DaEun Choi, Yoonseo Choi, and Juho Kim. 2022.
\newblock \href {https://doi.org/10.1145/3491102.3501931} {Stylette: Styling
  the web with natural language}.
\newblock In \emph{Proceedings of the 2022 CHI Conference on Human Factors in
  Computing Systems}, CHI '22, New York, NY, USA. Association for Computing
  Machinery.

\bibitem[{Laput et~al.(2013)Laput, Dontcheva, Wilensky, Chang, Agarwala,
  Linder, and Adar}]{laput2013pixeltone}
Gierad~P Laput, Mira Dontcheva, Gregg Wilensky, Walter Chang, Aseem Agarwala,
  Jason Linder, and Eytan Adar. 2013.
\newblock Pixeltone: A multimodal interface for image editing.
\newblock In \emph{Proceedings of the SIGCHI Conference on Human Factors in
  Computing Systems}, pages 2185--2194.

\bibitem[{Li and Li(2022)}]{li2022spotlight}
Gang Li and Yang Li. 2022.
\newblock Spotlight: Mobile ui understanding using vision-language models with
  a focus.
\newblock \emph{arXiv preprint arXiv:2209.14927}.

\bibitem[{Li et~al.(2020)Li, He, Zhou, Zhang, and
  Baldridge}]{li-etal-2020-mapping}
Yang Li, Jiacong He, Xin Zhou, Yuan Zhang, and Jason Baldridge. 2020.
\newblock \href {https://doi.org/10.18653/v1/2020.acl-main.729} {Mapping
  natural language instructions to mobile {UI} action sequences}.
\newblock In \emph{Proceedings of the 58th Annual Meeting of the Association
  for Computational Linguistics}, pages 8198--8210, Online. Association for
  Computational Linguistics.

\bibitem[{Li et~al.(2021)Li, Li, Zhou, Dehghani, and Gritsenko}]{li2021vut}
Yang Li, Gang Li, Xin Zhou, Mostafa Dehghani, and Alexey Gritsenko. 2021.
\newblock Vut: Versatile ui transformer for multi-modal multi-task user
  interface modeling.
\newblock \emph{arXiv preprint arXiv:2112.05692}.

\bibitem[{Ljungholm()}]{ljungholmvoice}
Alice Ljungholm.
\newblock Voice interaction vs screen interaction when controlling your
  music-system.
\newblock In \emph{CONFERENCE IN INTERACTION TECHNOLOGY AND DESIGN}, page 103.

\bibitem[{Luger and Sellen(2016)}]{luger2016like}
Ewa Luger and Abigail Sellen. 2016.
\newblock " like having a really bad pa" the gulf between user expectation and
  experience of conversational agents.
\newblock In \emph{Proceedings of the 2016 CHI conference on human factors in
  computing systems}, pages 5286--5297.

\bibitem[{Mao et~al.(2016)Mao, Huang, Toshev, Camburu, Yuille, and
  Murphy}]{mao2016generation}
Junhua Mao, Jonathan Huang, Alexander Toshev, Oana Camburu, Alan~L Yuille, and
  Kevin Murphy. 2016.
\newblock Generation and comprehension of unambiguous object descriptions.
\newblock In \emph{Proceedings of the IEEE conference on computer vision and
  pattern recognition}, pages 11--20.

\bibitem[{Mardanbegi and Hansen(2011)}]{mardanbegi2011mobile}
Diako Mardanbegi and Dan~Witzner Hansen. 2011.
\newblock Mobile gaze-based screen interaction in 3d environments.
\newblock In \emph{Proceedings of the 1st conference on novel gaze-controlled
  applications}, pages 1--4.

\bibitem[{Mayer et~al.(2020)Mayer, Laput, and Harrison}]{mayer2020}
Sven Mayer, Gierad Laput, and Chris Harrison. 2020.
\newblock \href {https://doi.org/10.1145/3313831.3376479} {Enhancing mobile
  voice assistants with worldgaze}.
\newblock In \emph{Proceedings of the 2020 CHI Conference on Human Factors in
  Computing Systems}, CHI '20, page 1–10, New York, NY, USA. Association for
  Computing Machinery.

\bibitem[{Pasupat et~al.(2018)Pasupat, Jiang, Liu, Guu, and
  Liang}]{pasupat-etal-2018-mapping}
Panupong Pasupat, Tian-Shun Jiang, Evan Liu, Kelvin Guu, and Percy Liang. 2018.
\newblock \href {https://doi.org/10.18653/v1/D18-1540} {Mapping natural
  language commands to web elements}.
\newblock In \emph{Proceedings of the 2018 Conference on Empirical Methods in
  Natural Language Processing}, pages 4970--4976, Brussels, Belgium.
  Association for Computational Linguistics.

\bibitem[{Rong et~al.(2017)Rong, Yi, and Tian}]{rong2017unambiguous}
Xuejian Rong, Chucai Yi, and Yingli Tian. 2017.
\newblock Unambiguous text localization and retrieval for cluttered scenes.
\newblock In \emph{Proceedings of the IEEE Conference on Computer Vision and
  Pattern Recognition}, pages 5494--5502.

\bibitem[{Rong et~al.(2019)Rong, Yi, and Tian}]{rong2019unambiguous}
Xuejian Rong, Chucai Yi, and Yingli Tian. 2019.
\newblock Unambiguous scene text segmentation with referring expression
  comprehension.
\newblock \emph{IEEE Transactions on Image Processing}, 29:591--601.

\bibitem[{Rozanova et~al.(2021)Rozanova, Ferreira, Dubba, Cheng, Zhang, and
  Freitas}]{rozanova2021grounding}
Julia Rozanova, Deborah Ferreira, Krishna Dubba, Weiwei Cheng, Dell Zhang, and
  Andre Freitas. 2021.
\newblock Grounding natural language instructions: Can large language models
  capture spatial information?
\newblock \emph{arXiv preprint arXiv:2109.08634}.

\bibitem[{Singh et~al.(2019)Singh, Natarajan, Shah, Jiang, Chen, Batra, Parikh,
  and Rohrbach}]{singh2019towards}
Amanpreet Singh, Vivek Natarajan, Meet Shah, Yu~Jiang, Xinlei Chen, Dhruv
  Batra, Devi Parikh, and Marcus Rohrbach. 2019.
\newblock Towards vqa models that can read.
\newblock In \emph{Proceedings of the IEEE/CVF Conference on Computer Vision
  and Pattern Recognition}, pages 8317--8326.

\bibitem[{Vtyurina et~al.(2019)Vtyurina, Fourney, Morris, Findlater, and
  White}]{vtyurina2019}
Alexandra Vtyurina, Adam Fourney, Meredith~Ringel Morris, Leah Findlater, and
  Ryen~W. White. 2019.
\newblock \href {https://doi.org/10.1145/3308558.3314136} {Bridging screen
  readers and voice assistants for enhanced eyes-free web search}.
\newblock In \emph{The World Wide Web Conference}, WWW '19, page 3590–3594,
  New York, NY, USA. Association for Computing Machinery.

\bibitem[{Wang et~al.(2022)Wang, Li, and Li}]{wang2022enabling}
Bryan Wang, Gang Li, and Yang Li. 2022.
\newblock Enabling conversational interaction with mobile ui using large
  language models.
\newblock \emph{arXiv preprint arXiv:2209.08655}.

\bibitem[{Xu et~al.(2021)Xu, Masling, Du, Campagna, Heck, Landay, and
  Lam}]{xu-etal-2021-grounding}
Nancy Xu, Sam Masling, Michael Du, Giovanni Campagna, Larry Heck, James Landay,
  and Monica Lam. 2021.
\newblock \href {https://doi.org/10.18653/v1/2021.naacl-main.80} {Grounding
  open-domain instructions to automate web support tasks}.
\newblock In \emph{Proceedings of the 2021 Conference of the North American
  Chapter of the Association for Computational Linguistics: Human Language
  Technologies}, pages 1022--1032, Online. Association for Computational
  Linguistics.

\bibitem[{Yu et~al.(2018)Yu, Lin, Shen, Yang, Lu, Bansal, and
  Berg}]{yu2018mattnet}
Licheng Yu, Zhe Lin, Xiaohui Shen, Jimei Yang, Xin Lu, Mohit Bansal, and
  Tamara~L Berg. 2018.
\newblock Mattnet: Modular attention network for referring expression
  comprehension.
\newblock In \emph{Proceedings of the IEEE Conference on Computer Vision and
  Pattern Recognition}, pages 1307--1315.

\bibitem[{Zhang et~al.(2021)Zhang, de~Greef, Swearngin, White, Murray, Yu,
  Shan, Nichols, Wu, Fleizach, Everitt, and Bigham}]{zhang2021screen}
Xiaoyi Zhang, Lilian de~Greef, Amanda Swearngin, Samuel White, Kyle Murray,
  Lisa Yu, Qi~Shan, Jeffrey Nichols, Jason Wu, Chris Fleizach, Aaron Everitt,
  and Jeffrey~P Bigham. 2021.
\newblock \href {https://doi.org/10.1145/3411764.3445186} {Screen recognition:
  Creating accessibility metadata for mobile applications from pixels}.
\newblock In \emph{Proceedings of the 2021 CHI Conference on Human Factors in
  Computing Systems}, CHI '21, New York, NY, USA. Association for Computing
  Machinery.

\end{thebibliography}
\bibliographystyle{acl_natbib}

\appendix

\section{Appendix}
\label{sec:appendix}

\subsection{Annotation Guidelines for Category Level Request Collection}
\label{category_guidelines}
In this project, you will be shown an entity category (phone number, url etc). Assume you see a particular instance of that entity on your screen. You have to come up with various requests you would say to a Voice Assistant to perform action on that. 
The main idea is to provide varied natural ways of interacting with that entity. The request should be one which holds valid when looking at different kinds of images containing that entity. 
 
Consider you see that entity on the screen. \textbf{Do not} assume any other information about that entity,  like what digits occur in the number or what place is the address for (note to readers - such references are the focus of the unambiguous request collection, hence skipped here).

\begin{itemize}
    \item Take me to the California address - Incorrect
    \item Call the number ending in 99 - Incorrect
    \item Take me to COUNTRYNAME address - Incorrect
    \item  Call COMPANYNAME number - Incorrect
\end{itemize}

\begin{enumerate}
    \item You are encouraged to use varied request formulations with different ways of referring to the entity as well as carrying out different actions on that entity. Example - Phone number
    \begin{itemize}
        \item place a call to that phone number
        \item dial this number
        \item add this to my contacts
        \item remind me to call here at 5
        \item send this to PERSON on text message
    \end{itemize}
    \item  Constraints to follow -
    \begin{enumerate}
        \item Enter requests only in column 1 and do not change the values in column 2 in any way.
        \item The number in the second column reflects the number of unique requests so far. 
        \begin{enumerate}
            \item When you enter a request and the number does not increase, this means that request is already present. CHANGE the request.
            \item When you enter a request and the number in column 2 increases by 1, you have successfully entered a unique request. Move to the next row.

        \end{enumerate}
    \item Do not make minor irrelevant variations.  Replace proper nouns with uppercase tags like “PERSON”, “COMPANYNAME”, “DAY”.
    \begin{enumerate}
        \item Incorrect - 
        \begin{itemize}
            \item send this to Mom on text message 
            \item send this to Dad on text message 
            \item send this to John on text message 
        \end{itemize}
        \item Correct - 
        \begin{itemize}
            \item send this to PERSON on text message
            \item share the number with PERSON
        \end{itemize}
        
    \end{enumerate}
    \item Use \textbf{only lowercase letters} in the request, apart from the proper noun replacements with all uppercase tags (PERSON, COMPANYNAME etc). Use these only in a way that one can replace them with any name without knowing the actual screen. The request should hold valid for a variety of different screens containing phone numbers.
    \begin{enumerate}
        \item Incorrect - Send this to PERSONNAME on text message - first letter should be small
        \item Incorrect - copy PERSONNAME’s number - assumes you see PERSONNAME
        \item Correct - send this to PERSONNAME on text message 
        \item Correct - send PERSONNAME’S number to this number - here PERSONNAME can be any person in your contacts.
    \end{enumerate}
    \item No fullstops after the request.
        1. Incorrect - call this.
        2. Correct - call this
    \item No trailing or leading whitespaces should be added.
    \item  Assume you see the entity in front of you. Target the request to ask a VA to \textit{act on the entity type mentioned}. Do not just add the entity type in the request randomly. Do not assume anything more about what you see.
    Invalid requests - 
    \begin{itemize}
        \item tell me the number that just called me - the request is not about a phone number you are seeing - “did I just receive a call from that number” is a valid request
        \item is PERSON's number in my missed calls - the request is not about a number you are seeing - 
                “is this number in my missed calls” is valid
                \item get rid of COMPANYNAME's number - You may not be seeing a company name- get rid of their number is valid
    \end{itemize}
            \item  Use varied ways to refer to the entities. For instance, for `phone number', You can use generic references like “this”,  “that” as well as phrases including “phone number”, “contact number” etc.
            \begin{itemize}
                \item call that 
        \item  call this contact number
        \item call them
        \item call this number
            \end{itemize}
    \item You need not explicitly use the phrase mentioning the entity type always, specially if the intent conveys that. Example -  Email address
        \begin{itemize}
            \item draft a mail to this
            \item  draft them a mail
        \end{itemize}
    \item Use \textit{varied ways of referring} to the entity
        \begin{itemize} 
        \item generic phrases - this/that/it/them/.... etc
        \item specific phrases - 
            email/email address/address/contact/... etc
            \end{itemize}
    \end{enumerate}

\end{enumerate}
\begin{figure}[hbt]
     \centering
     \begin{subfigure}[b]{0.49\textwidth}
        \centering
\includegraphics[width=0.95\textwidth]{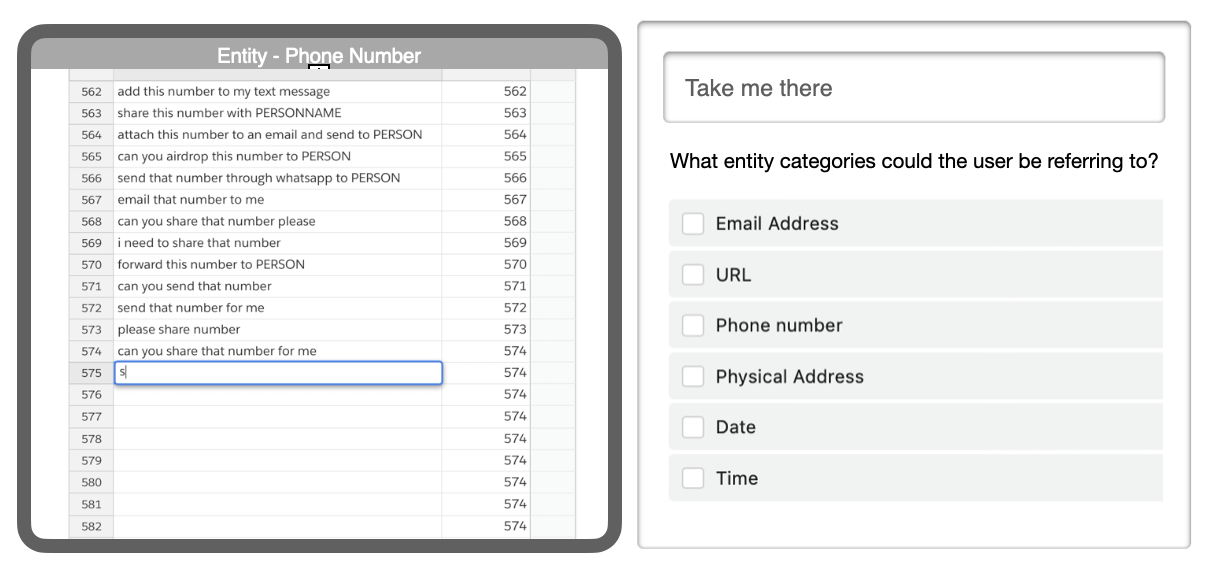}
        \caption{\textbf{Category Level Data Collection}: First, annotators are asked to provide category level requests in a spread sheet (Left). Column2 of the sheet reflects the unique count so far, which encourages varied requests for a diverse dataset. We define constraints in the guidelines so that variations are not spurious changes. Second, annotators are asked to verify the collected requests to capture entity level ambiguity (Right). 3 annotators are asked to verify each collected request.}
        \label{fig:cat_collection}
     \end{subfigure}
     \hfill
          \begin{subfigure}[b]{0.49\textwidth}
        \centering
        \includegraphics[width=0.95\textwidth]{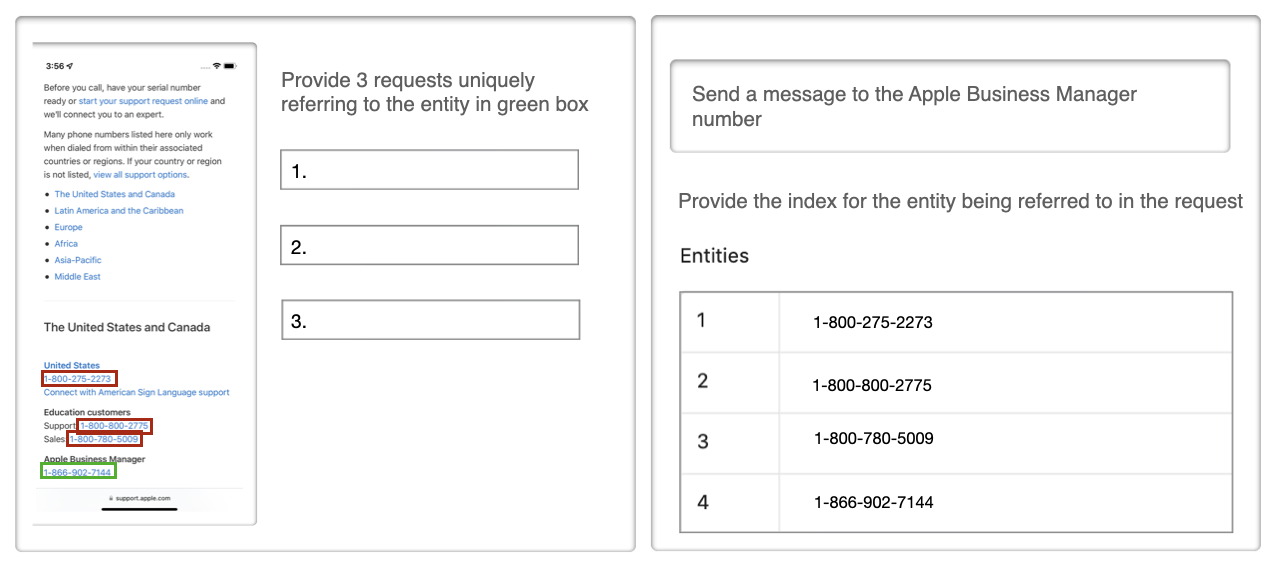}
        \caption{\textbf{Unambiguous Data Collection}: First, annotators are shown a screen with multiple instances of a category (Left). One is highlighted in a green box, while others in red boxes as initially annotators tended to provide ambiguous references.  An annotator provides 3 different requests with references. Second, the correctness of a request is verified by showing the screen and request (Right). 3 annotators are asked to mark the referred one.}
        \label{fig:unamb_collection}
     \end{subfigure}
        \caption{Unambiguous and Category level Data collections protocols.}
        \label{fig:data_collection}
\end{figure}
\subsection{Annotation Guidelines for Unambiguous Request Collection}

\label{descriptive_guidelines}
\textbf{Overview}
The goal of this task is to generate a variety of requests for text in a screen. The requests should be queries or requests you would make to a voice assistant, based on the text. 
You will be shown a screen with a green bounding box around specific text. You will need to:
\begin{quote}
    Write three uniquely referential requests about the marked text for a voice assistant
\end{quote}

\subsubsection{Green vs. Red Boxes}
Screens will contain green and red boxes. 
The green box contains the text for which you need to write the requests. The requests for the text within the green box need to uniquely identify it.
Red boxes mark the texts that are similar to the text within the green box. For example, if an image has three phone numbers, the red box will capture the other two phone numbers. Do not write requests for the text within the red box. They are intended to serve as a guidance so that you don't miss them out and ensure you write uniquely referential requests for the text within the green box. 

\subsubsection{Request Guidelines}
Imagine you are viewing that screen on your phone, and were to ask a voice assistant about that text you came across.
What would you ask the voice assistant regarding the text that you could not gather just from looking at it? 
What additional actions or requests would you ask the voice assistant to execute in relation to the text that can be carried out on your mobile device?

Keep in mind the following:
\begin{itemize}
\item Unique: Each request will require a referring expression that uniquely identifies the detected text.
\item Require a voice assistant's help: Requests should not be questions that a user can answer simply by looking at the text. 
Example: “Does this phone number contain 007 at the end” is invalid.
\item Mix it up: Requests can be a mix of questions about the text or action commands to be executed on the text. 
\item Sound natural: Come up with requests that would sound natural, coming from a user. Verbally say the request out loud to ensure it sounds natural and not too long.
\item Make sense for a user to request a VA: Think about whether the request would make sense for a user to request, based upon the text type/context of the screen, and what a user would usually do on a device with that information.
\end{itemize}

\textbf{Uniquely Referential}
Use references that ensure the request uniquely identifies the marked text. All 3 requests for a particular text need to be uniquely referential. Use varied actions and request types. Do not use the same reference across the requests.
Remember that the request needs to be uniquely referential, not just with other similar texts marked in red, but also with all content within the screen.
Example \textbf{errors}:
\begin{itemize}
    \item Too General:
\begin{enumerate}
    \item Call that
    \item Text it to John
    \item Save that to my notes
\end{enumerate}
\item Same references for all 3 requests:
\begin{enumerate}
    \item Call the third number
    \item Share the third number
    \item Copy the third number to my notes
\end{enumerate}
\end{itemize}

\end{document}